# Dwarf spheroidals and the evolution of not-too-old Population II stars.

V. Castellani[1] and S. Degl'Innocenti[2]

[1] *Dipartimento di Fisica dell'Universita' di Pisa 50100 Pisa, Italy and Osservatorio Astronomico di Collurania Teramo, Italy*

[2] *Dipartimento di Fisica universita' di Ferrara and INFN sezione di Ferrara 44100 Ferrara, Italy*

**Abstract**

Previous evolutionary computations concerning stars with masses in the range of the "Red Giants Transition Phase" are extended to lower metallicities. Clusters isochrones for two values of the metallicity Z ($Z=10^{-4}$, $4\cdot10^{-4}$) and for ages down to 1 billion years are presented, discussing selected evolutionary features characterizing the structure of the stars during both H or He burning phases. One finds that clusters with $Z\simeq 10^{-4}$ and with an age around 1-2 billion years could be characterized by the occurrence of anomalous He burning variables at a luminosity of the order of LogL= 2.0-2.2 $L_{\odot}$.

## 1 Introduction

According to a well established observational scenario, one recognizes in the Galaxy two main families of stars, as given by old, metal poor stars membering the halo (Population II) or younger, metal rich, solar-like stars membering the galactic disk (Population I). Correspondingly, a great deal of theoretical effort has been devoted in the literature investigate this kind of stellar structures. However, the evidence has grown that such a distribution of stellar parameters is far from being exhaustive of the stellar phenomenon in the Universe. When looking far away from our Galaxy one finds stars which do not fit such a simple, schematic classification. Evidences for young, metal poor stellar clusters (like NGC 330) have been already found in the Magellanic Clouds. On the other side, elliptical galaxies provide us with the convincing evidence for old, metal rich, stellar populations. Thus the original concept of "Population" is not more adequate to cover the reality of actual stars in the Universe.

According to previous discussions (see, e.g., Castellani 1986) let us keep the concept of "Population" as only connected with the star metallicity. That is, let us privilege the "age" of the matter with respect to the age of the stars. Thus, metal poor stars, independently of their age, should be regarded as "Population II" stars, since born out from matter only little evolved from the primeval Big-Bang mixture. Since the amount of heavy element in a star is a key parameter governing the evolution of stellar structure, one can regard the previous definition as a reasonable way to keep in a same class stars with similar evolutionary behavior. Thus let us speak about old or young population II stars.

In this sense, one finds that the two extreme ages for Pop II stars have been rather extensively explored in the literature, in connection with the already quoted evidence from the Galaxy and from the nearby Magellanic clouds. However the evidence is increasing for dwarfs spheroidals being true old Population II systems, but not-too-old., i.e., not as old as the galactic globulars whose age is believed of the order of the Hubble time. As an example one can recall the case of the spheroidal Leo I, for which Lee et al. (1993) suggest an age of the order of 3 Gyr. According to this suggestion Leo I would be sensitively younger than a typical galactic old disk cluster as the well known NGC 188 which is attributed of, about, 6 Gyr.

Similar evidences suggest the opportunity to extend to very low metallicities evolutionary computations already presented to connect the range of low mass stars to that of intermediate mass through the so called "Red Giant Transition Phase" (RGTP: Sweigart, Greggio & Renzini 1989, Castellani, Chieffi & Straniero 1990). This paper is devoted to present the results of such an investigation. In the following section we will investigate the Hydrogen burning evolution for a selected range of stellar masses as computed for suitable choices on the star metallicity. On this basis section 3 will deal with the results concerning He burning phase, discussing some interesting connections with observations. A final discussion will close the paper.

## 2 H-burning phases

As quoted in the introduction, evolutionary evidences from old galactic globulars already drove an impressive amount of theoretical work on the corresponding evolutionary scenario. As a consequence, the H burning evolution of metal poor, low mass stars ($0.6 \leq M \leq 1.0\,M_\odot$) is among the best established results of current evolutionary theories. The extension of this scenario to larger masses has been however produced only for $Z \geq 4 \cdot 10^{-3}$ (Sweigart, Greggio & Renzini 1989) and, more recently, for $Z \geq 10^{-3}$ (Castellani, Chieffi & Straniero 1990: CCS). According to such an occurrence, in the following we will discuss the evolutionary behavior of stars up to, at least, M=2 $M_\odot$ for the two selected choice on the amount of heavy elements $Z=10^{-4}$, $4{\cdot}10^{-4}$. To this purpose, evolutionary tracks already presented by Cassisi and Castellani (1993) for $Z=10^{-4}$ have been implemented with new tracks for suitable choices of the stellar masses, adding a set of new computations for the case $Z=4 \cdot 10^{-4}$. In all case the original He content has been taken fixed at Y=0.23, i.e., at the value which is believed to be adequate for these population II stars. On this basis, we will able to extend available computations to cluster ages lower than 1 Gyr.

As already known, all stars in the quoted mass range do undergo a violent ignition of the central He burning through the so called He-flash. All the models have been thus followed till the onset of the He flash, in order to obtain the two key parameters needed to evaluate the following He burning phases, namely the mass of the He core at the flash ($M_c$) and the amount of extrahelium brought to the surface by the first dredge up ($\Delta Y$). Table 1 gives the list of these parameters when $Z=10^{-4}$ for all the computed evolutionary sequences, together with the values for both the age and the stellar luminosity at the flash. Note the fine grid of masses computed for this metallicity beyond M=2 $M_\odot$ to gain detailed information about the behavior of stellar structures through the RGTP. Table 2 gives similar data but for $Z=4{\cdot}10^{-4}$.

Selected isochrones, as computed on the basis of the referred computations, are presented in Fig. 1, for the two different assumptions about the star metallicity. Both evolutionary tracks and isochrones are available by E-mail upon request. As for the luminosity of the turn-off, which is the well known key parameter to evaluate the cluster age, by interpolation of our results one gets:

$$\mathrm{LogL}_{TO} = -1.01\ \mathrm{Logt}_9 + 1.56 \qquad Z=10^{-4}$$
$$\mathrm{LogL}_{TO} = -0.96\mathrm{Logt}_9 + 1.46 \qquad Z=4 \cdot 10^{-4}$$

which reproduces evolutionary results within a few hundreds of magnitude when the isochrones are not affected by evidence of overall contraction phase, and which appear in good agreement with the evaluations presented by Straniero & Chieffi (1991). When the overall contraction appears the previous relations keep giving with rather good accuracy the luminosity of the stars just beyond the end of the overall contraction. As a whole, one finds that the (linear) dependence of the TO luminosity on $\log t_9$ derived for globular clusters can be safely extrapolated toward reasonably lower ages.

As already known, this is not the case for both $M_c$ and $\Delta Y$ which deserves some further comments. Fig. 2 discloses the behavior of $M_c$ as a function of the mass of the evolving star for the case $Z=10^{-4}$. One can follow in details the progressive decrease of $M_c$ when the stellar mass increases and the electronic degeneracy is progressively removed. One finds that the minimum value of $M_c$ is attained when $M \sim 2.25\ M_\odot$. For larger masses $M_c$ starts to rise again, for the very simple reason that stars during the MS phase begin to experience an increasing central convective core, which now governs the depletion of H in the stellar interior.

According to the discussion given in CCS, one can assume as "critical mass" the less massive star experiencing a central flash, i.e. $M \simeq 1.7\ M_\odot$, which, in turn, corresponds to the model roughly half way in the core decrease. According to previous studies, one knows that this critical mass is increasing with metallicity. Comparison with data for $Z=4 \cdot 10^{-4}$, as given in the same fig. 2, confirms this occurrence, showing in the same time that at the lower mass limit $M_c$ decreases when metallicity is increased, as well known since the pioneering paper by Sweigart and Gross (1978). Fig. 2 shows that increasing the metallicity the degeneration is removed with more difficulty, reconciling the smaller core mass of low mass,lower metallicity stars with the larger value for their critical mass.

Fig. 3 shows the amount of extrahelium brought to the surface of our models by the first dredge up. Comparison with previous results taken from the literature, as given in the same figure, discloses a general agreement. One recognizes that in all cases the model with $M= 1.3\ M_\odot$ marks the separation between two regimes: the regime of low mass stars where the extrahelium increases when the stellar mass or the metallicity is increased, as early recognized by Sweigart & Gross (1978), and the regime of RGTP where just the contrary holds, as found by Sweigart, Greggio & Renzini (SGR: 1989) These evidences have to be discussed in connection with the prescriptions about efficiency of surface convection. According to these prescriptions, external convection sinks deeper when the stellar mass decreases or metallicity increases. An occurrence well studied in the low mass regime, in connection with the luminosity of the RG bumps, and confirmed by SGR also for larger masses.

However, in spite of the deeper convection, in the low mass regime one finds that the efficiency of the dredge up is increasing with increasing the mass. Without enter into details, the reason is that, until the central H burning is supported by the proton-proton reactions only, the need for a larger luminosity push the burning more and more toward the exterior. This occurrence governs the efficiency of the dredge-up in spite of the opposite effect of convection. The opposite behavior in the RGTP regime is the signal that CNO burning is increasing its efficiency, concentrating toward the center the production of the energy and, in turn, of He.

On the basis of such an approach, now one can understand why, increasing Z, $\Delta$Y increases (decreases) below (above) M $\simeq$ 1.5 $M_{\odot}$, as shown in the same fig.3 and already found in SGR. Increasing Z the convection goes deeper in the structure, thus increasing the dredge-up. However, the CNO burning starts earlier decreasing the abundance of He in the convective zone, according to the previous discussions. Let us finally notice that the peculiar lack of extrahelium for M $\geq$ 2.5 $M_{\odot}$ is only the consequence of the fact that very metal poor stars above this mass fail reaching the Hayashi track before the ignition of He (see, e.g., Cassisi & Castellani 1993), thus skipping the dredge-up phase.

## 3 The He burning phase

Evolutionary data for H-burning stars, as given in the previous section, allow to us to investigate in some details the evolution of these stars during the following He burning phase. The procedure, in principle, is the one well experienced in investigating the HB location in galactic globulars. For each assumption about the cluster age and on its original chemical composition, one derives the mass ($M_{RG}$) of the star at the He flash and the value of both $M_c$ and $\Delta$Y for that star. ZAHB models have thus to be derived as the sequence of stellar models burning at the center of a core of mass $M_c$ surrounded by an envelope enriched by $\Delta$Y under the condition that the total mass of this structures ($M_c$ + $M_{envelope}$) must be smaller or equal to $M_{RG}$.

In our case, the only substantial difference will be that $M_c$ cannot more re-garded as largely independent of the cluster age. Thus, for each assumption on the cluster age one will have different values for $M_c$ and, in turn, different luminosities for these "HB" stars. In this way we produced theoretical expectations for the ZAHB structures corresponding to selected assumptions about the cluster ages. Note that this "evolutionary" procedure differs from similar computations early presented either by Demarque & Hirshfeld (1975) for a constant mass of the He core or by Hirshfeld (1980), who gave to He burning stars the mass of the He core (at the He flash) of a Red Giants with the same mass

Fig. 4 shows the HR diagram location of these ZAHB as computed for $Z=10^{-4}$ or $4\ 10^{-4}$ and for the labeled assumptions about the cluster ages. Inspection of this figure shows that, as already know (see, e.g. Caloi, Castellani & Tornambe 1978), in very metal poor cluster increasing the star mass of a ZAHB structure decreases -as normal- the surface temperature only to reach a minimum temperature after that the temperature starts increasing with mass. However, as can be observed in the same Fig. 4, such a behavior only becomes effective for cluster age lower or of the order of 2 Gyr, since for larger ages the evolutionary mass of RG stars is too small to fall on the "upper branch" which characterizes these very metal poor HBs.

The distribution of stars in Fig.4 becomes even more interesting when observing that some of these stars do fall in the instability strip for radial pulsations. The behavior depicted in Fig.4 can now be used to construct an interesting pulsational scenario, which fundamentally confirms the suggestions early presented in the quoted papers by Demarque & Hirshfeld (1975) and by Hirshfeld (1980). For the sake of the discussion, let us assume $Z= 10^{-4}$ and no mass loss. Since the instability strip can be fixed somewhere around logTe=3.8 - 3.9, young clusters cannot have variables, since the evolutionary paths of the larger masses is always hotter than the instability strip (see Cassisi and Castellani 1993). Increasing the cluster age toward 1 billion years, radial pulsators should appear, with a luminosity of the order of $\log L/L_\odot \simeq 2.0$. A further increase in age deprives again the cluster of variables, until at an age of about 15 billion years (RR) variables appear again, now at a luminosity $LogL \simeq 1.7$. A similar scenario can be obviously modified (and complicated) by relaxing the (unrealistic) hypothesis of no mass loss. As one can easily understand, by playing with the value of both the mean mass and the dispersion in masses one can produce various scenarios like the occurrence of underluminous RR-like variables at $LogL/L_\odot$ much lower than 1.7 and, in an extreme case, the contemporaneous occurrence of these variables with anomalous luminous variables at $LogL/L_{\odot} \simeq 2$. Discussing these scenario is out of the goals of this paper, which is only devoted to explore the field of theoretical possibilities.

Fig.4 shows how such a scenario is modified when passing from $Z=10^{-4}$ to $Z=4{\cdot}10^{-4}$. The "red edge" of the HB is now redder and the more massive stars can only marginally populate the instability region even for an age as low as $7 \cdot 10^8$ yr (and a mass as large as 2.0 $M_{\odot}$). Thus increasing the metallicity the previous scenario does vanish, and one can only expect more or less canonical RR Lyrae.

To investigate the evolutionary behavior of these "upper HB", the more massive models for both values of metallicities and for the various assumptions about the cluster age have evolved through the phase of central He burning till the exhaustion of central He, as shown in the same Fig. 4. One can see that the scenario produced on the basis of the ZAHB location is only little modified by evolution. The 1.7 $M_{\odot}$ ($Z=10^{-4}$) keeps evolving in the region of instability, though increasing its luminosity up to about $LogL/L_{\odot} \sim 2.2$. As a general rule, one finds that the blueward excursion shown by all the models can push within the strip some structure born near the red edge for instability, though not varying the qualitative scenario discussed above. Interesting enough, we finds that all these stars, though massive, have He burning evolutionary time of the same order of magnitude than canonical HB lifetimes, i.e., of the order of $10^8$ years. This allow the easy prediction of a consistent occurrence of He burning stars also in moderately young clusters.

As a final point, we show in Fig. 5 -as an example- the time behavior of the effective temperature for our 1.4 $M_{\odot}$, $Z=10^{-4}$ model. On this basis one expects to find these stars rather homogeneously distributed in temperature, though with a tendency to accumulate toward the two extreme temperatures. Data in Fig.5 indeed shows that in the case of our 1.4 $M_{\odot}$ model, the stellar population in the range of temperatures 3.75-3.77-3.79-3.81 is expected in the ratio 0.38:0.18:0.44.

## 4 Conclusions

This paper has been devoted to explore the evolutionary scenario for stars in not-too- old but very metal poor clusters. A preliminary investigation has been devoted to the evolution of H-burning structure. On this basis we extend available isochrones for very metal poor clusters to ages smaller than $\simeq$ 1 Gyr. The investigation of the following phases of central He burning gave the evidence that clusters with Z$\simeq 10^{-4}$ and with an age around 1-2 billion years could be characterized by the occurrence of anomalous variables, at a luminosity of the order of LogL= 2.0-2.2 $L_{\odot}$. Larger metallicities push the red limit of the HB toward smaller effective temperatures, preventing the further occurrence of such variables.

We present these computations as an useful implementation of the currently available evolutionary scenario, which will finds its natural field of application in studying the evolutionary status of some dwarf spheroidals (Caputo & Degl'Innocenti 1994) as well as, possibly, for some not-too-old globulars in the nearby Magellanic Clouds.

**Acknowledgment**

We thank Filippina Caputo who drove our attention on the problem, suggesting the need for such a completion of the current evolutionary scenario.

**References**

Caloi V., Castellani V. & Tornambe A. 1978 A&AS 33, 169

Caputo F. & Degl'Innocenti S. 1994, in preparation

Cassisi S. & Castellani V. 1993, ApJS 88, 509

Castellani V. 1986 Prog. Cosmic Phys. 9, 317

Castellani V., Chieffi A. & Straniero O., 1990, ApJ 74, 463

Chieffi A. & Straniero O. 1989, ApJS 71,47

Demarque P. & Hirshfeld A. 1975, ApJ 202,346

Hirshfeld A. 1980, ApJ 241,111

Lee M. G.,Freedman G.,Mateo M.,Thompson I.,Roth M., Ruiz M. 1990 AJ, 106, 1420

Straniero O. & Chieffi A. 1991, ApJS 76,525

Sweigart A., Greggio L. & Renzini A. 1989 ApJS 69
Sweigart A. & Gross P. 1978 ApJS 36,405

**Figure Captions**

Fig. 1: Selected Isochrones for the case $Z=10^{-4}$ (upper panel: t=0.7, 1, 2, 4, 7, 10 and 15 Gyr) and for $Z=4\cdot10^{-4}$ (lower panel: t= 1, 2, 4, 7, 10 Gyr).

Fig. 2: The behavior of the mass of the He-core at the He-flash as function of star masses for models with $Z=10^{-4}$ (full squares) or $Z=4\cdot10^{-4}$ (open squares)

Fig. 3: The amount of extra-helium brought to the surface by the first dredge-up for all the computed models. Open circles shows the results from Sweigart and Gross (1978; Y=0.20, Z= $10^{-4}$); crosses represent results from Chieffi & Straniero (1989: Y=0.23, Z= $2\cdot 10^{-4}$)

Fig. 4: The locus in the HR diagram of ZAHB structures for the various labeled assumptions about the cluster age when $Z=10^{-4}$ (upper panel) or $4\cdot10^{-4}$ (lower panel). For each ZAHB we give in the order: the cluster age, the mass of the more massive model, the surface He abundance, the mass of the He core at the beginning of the HB evolution. Masses are in solar units, ages in billion years. A lower mass limit of 0.7 $M_\odot$ has been assumed for all the ZAHB. Intervals by 0.1 $M_\odot$ are marked along the various ZAHB. The evolutionary path of the more massive model is also shown at the top of each ZAHB.

Fig. 5: Time behavior of the effective temperature for our 1.4 $M_\odot$, $Z=10^{-4}$ model.

Table 1: Selected evolutionary parameters at the He flash for Z=$10^{-4}$ (see text). Ages are in Gyr.

| $M/M_{\odot}$ | $t^{flash}$ | $L^{flash}$ | $\Delta Y$ | $M_c$ |
|---|---|---|---|---|
| 0.6 | 44 | 3.25 | 0.001 | 0.506 |
| 0.8 | 15 | 3.25 | 0.008 | 0.504 |
| 1.0 | 6.9 | 3.21 | 0.015 | 0.495 |
| 1.2 | 3.7 | 3.13 | 0.021 | 0.482 |
| 1.4 | 2.2 | 3.06 | 0.024 | 0.469 |
| 1.7 | 1.0 | 2.88 | 0.023 | 0.441 |
| 2.0 | 0.72 | 2.67 | 0.022 | 0.394 |
| 2.2 | 0.53 | 2.43 | 0.019 | 0.363 |
| 2.3 | 0.46 | 2.12 | 0.002 | 0.335 |
| 2.5 | 0.365 | 2.11 | 0.000 | 0.347 |

Table 2: Selected evolutionary parameters at the He flash for Z=$4{\cdot}10^{-4}$. Ages are in Gyr.

| $M/M_{\odot}$ | $t^{flash}$ | $L^{flash}$ | $\Delta Y$ | $M_c$ |
|---|---|---|---|---|
| 0.75 | 15 | 3.29 | 0.011 | 0.498 |
| 1.0 | 7.1 | 3.27 | 0.018 | 0.491 |
| 1.4 | 2.3 | 3.20 | 0.023 | 0.477 |
| 1.7 | 1.2 | 3.15 | 0.020 | 0.467 |
| 2.0 | 0.74 | 2.99 | 0.017 | 0.443 |

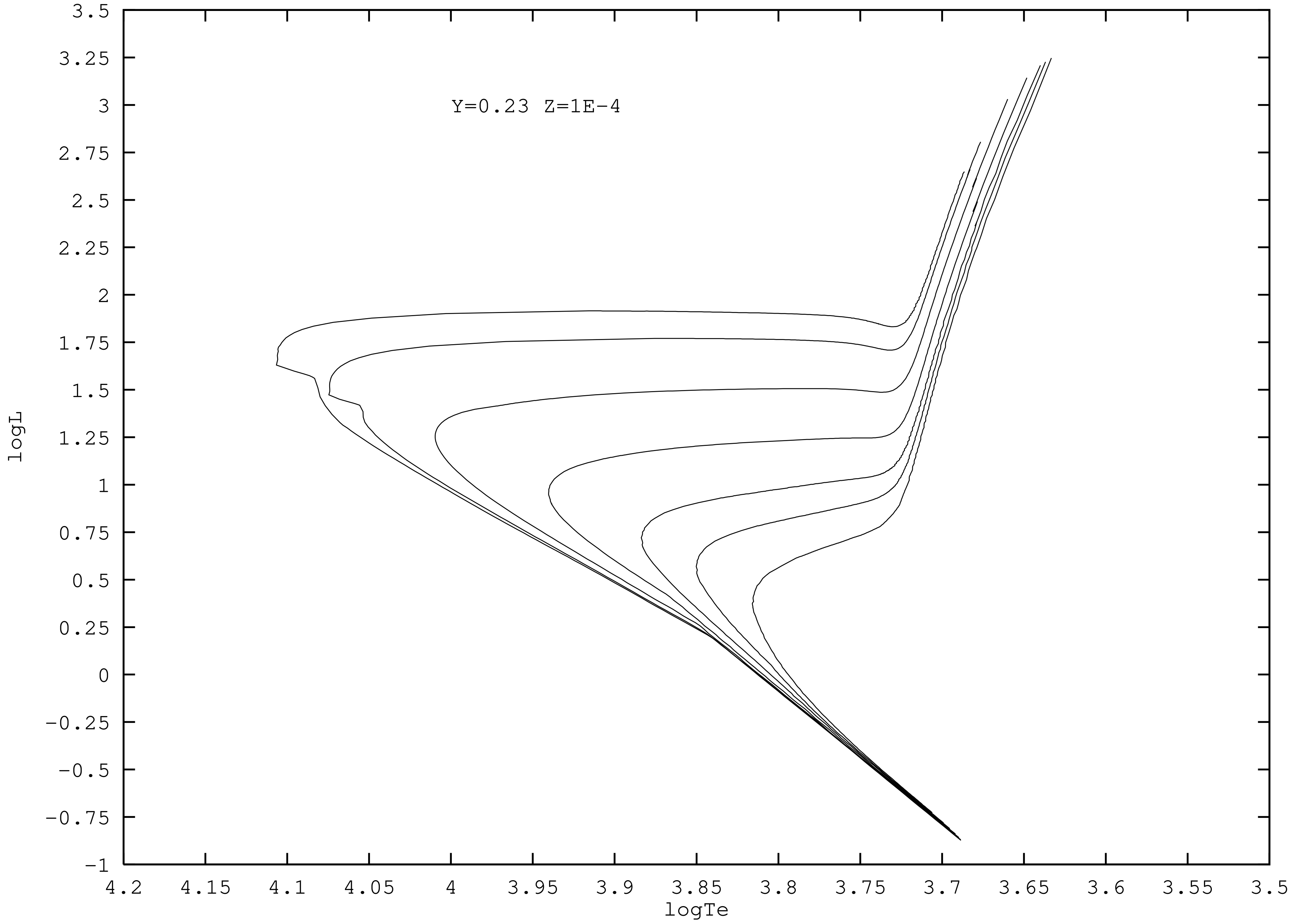

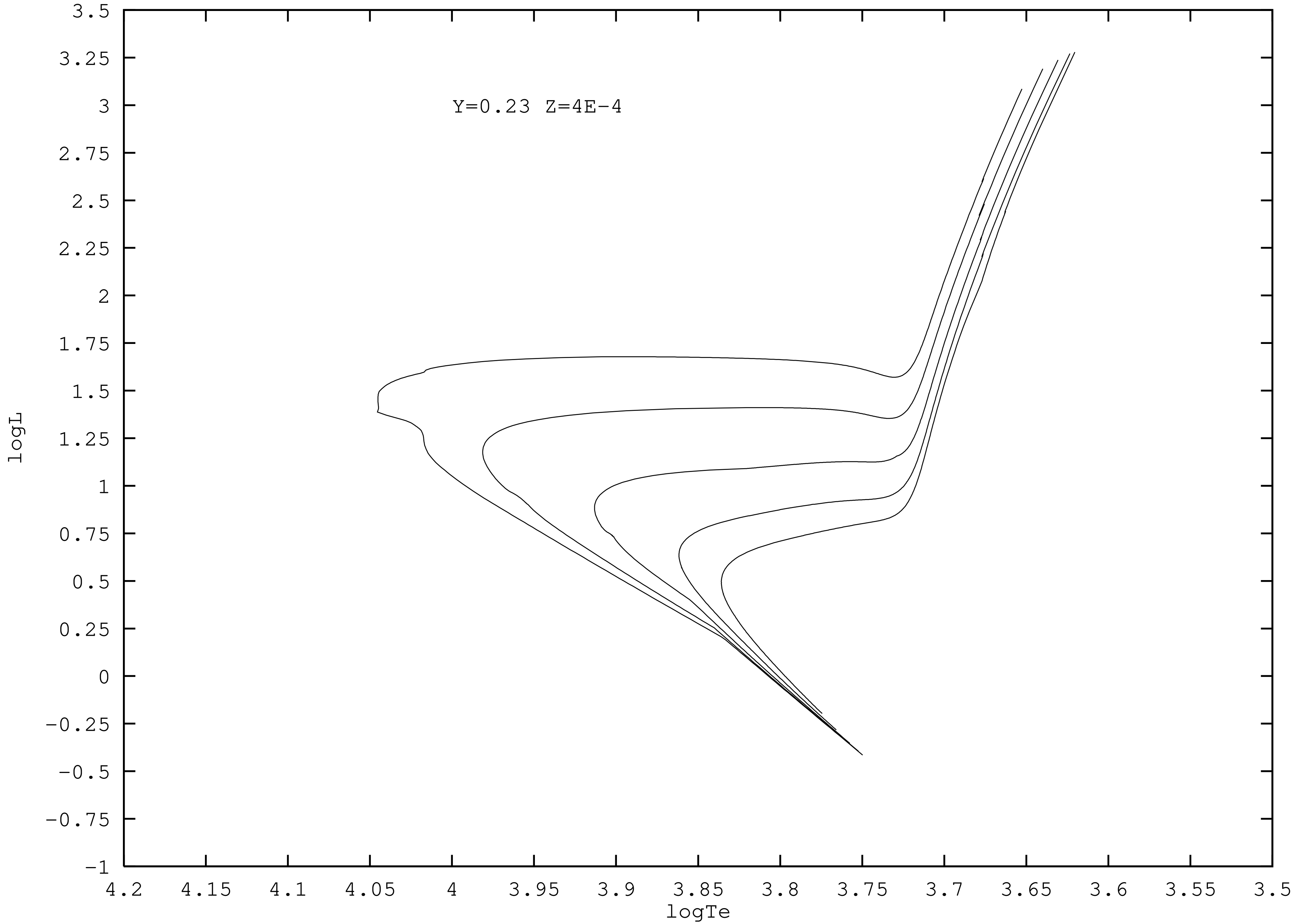

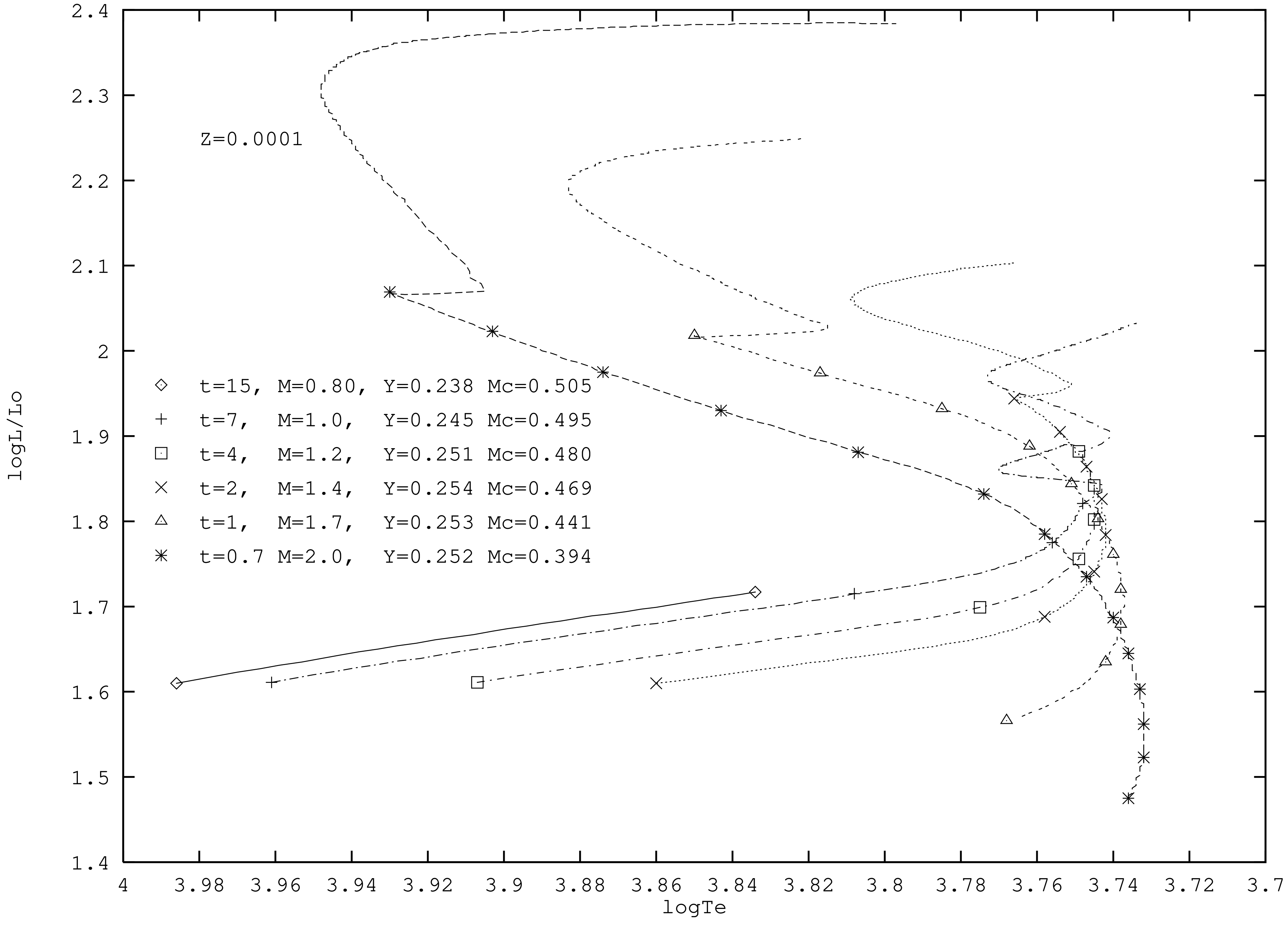

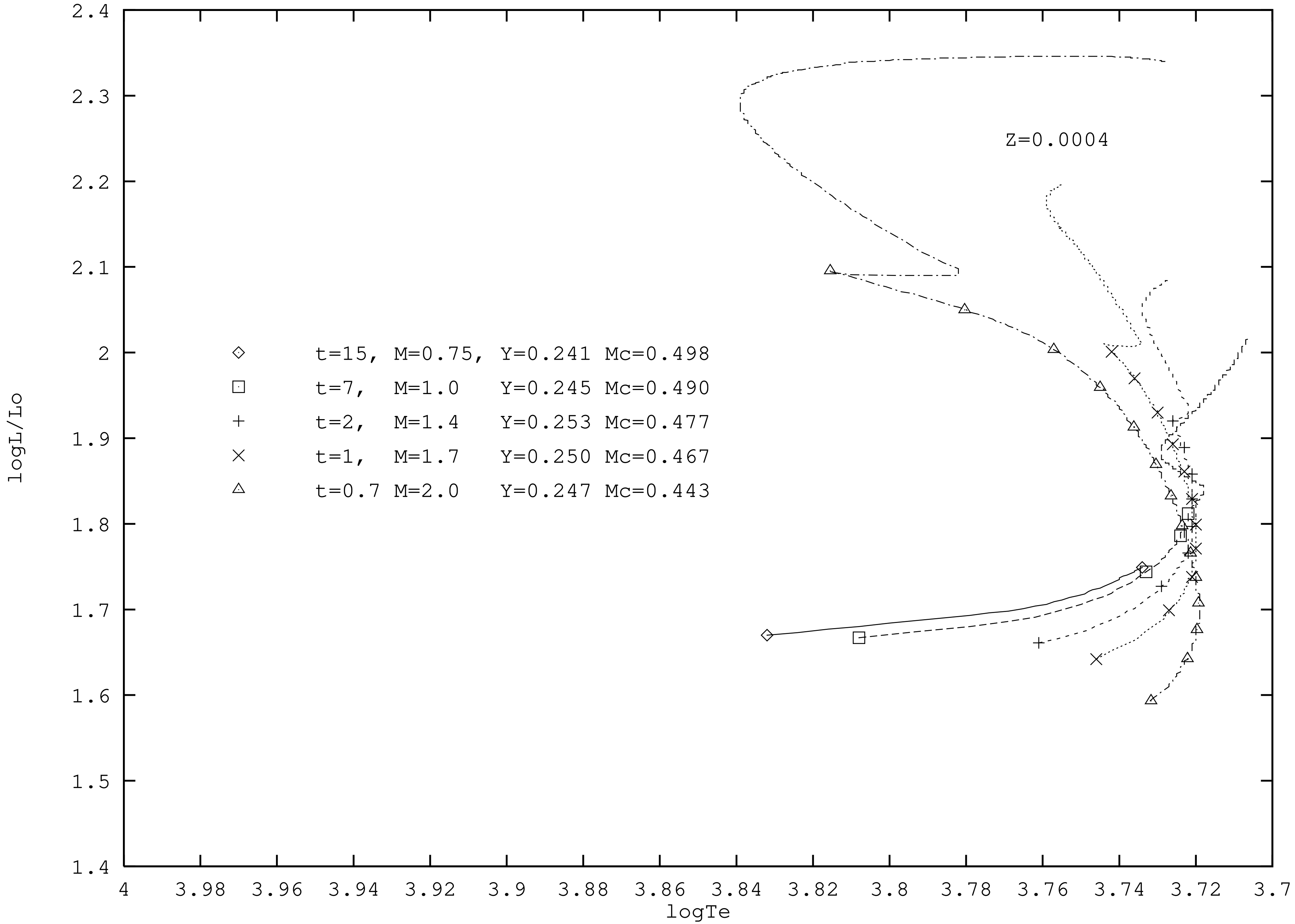

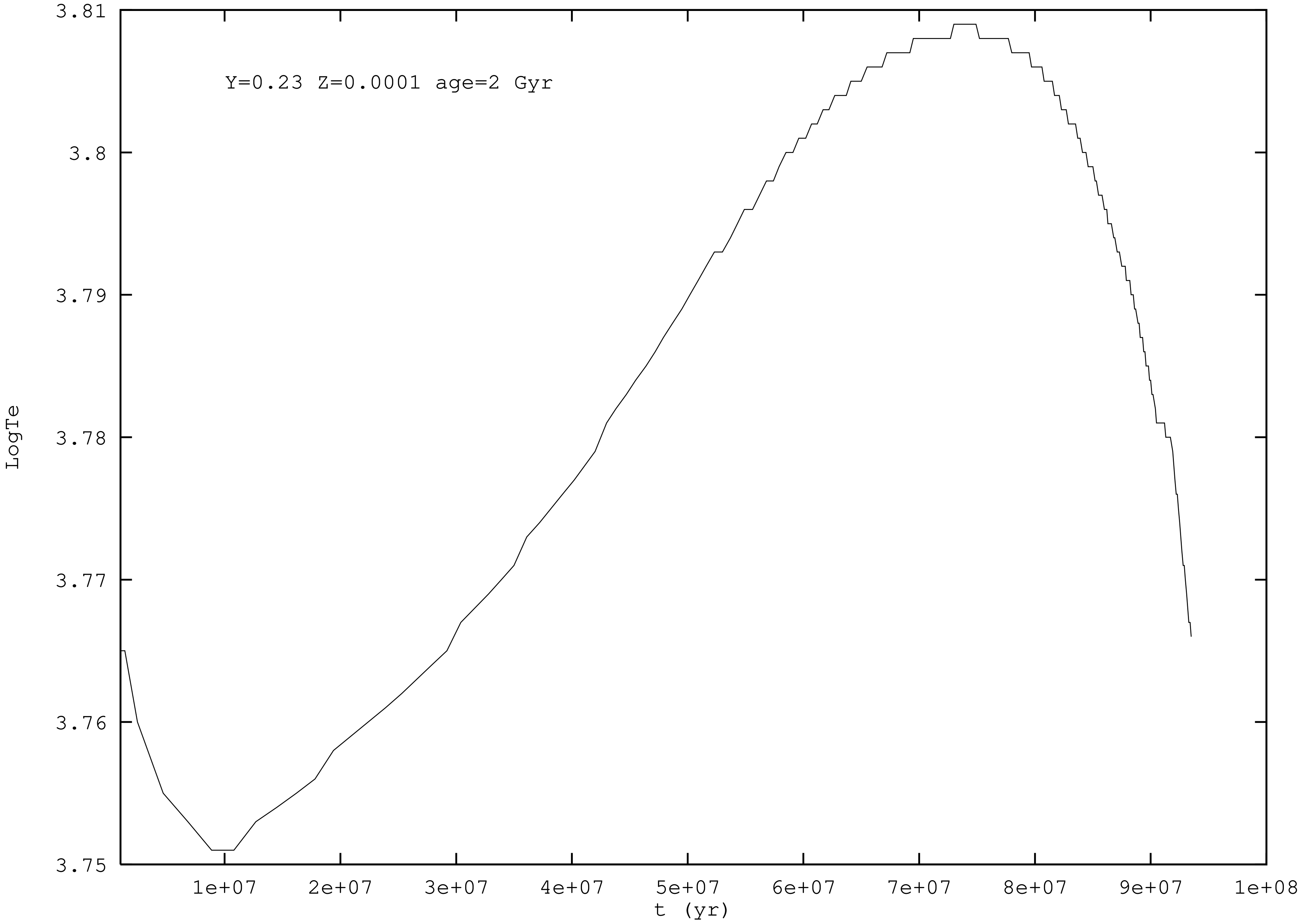